\documentclass[prd,floats,aps,showpacs]{revtex4}

\usepackage{dcolumn,epsfig}

\begin{document}

\title{Teukolsky evolution of particle orbits around Kerr black holes in the time domain: elliptic and inclined orbits}

\author{Gaurav Khanna}

\affiliation{Physics Department,
University of Massachusetts at Dartmouth, N. Dartmouth, MA 02747 }
\affiliation{Natural Science Division, Southampton College of Long Island University, Southampton NY 11968 }

\begin{abstract}
We extend the treatment of the problem of the gravitational waves produced by a particle of negligible mass orbiting a Kerr black hole using black hole perturbation theory \cite{Ramon} in the time domain, to elliptic and inclined orbits. We model the particle by smearing the singularities in the source term using narrow Gaussian distributions. We compare  results (energy and angular momentum fluxes) for such orbits with those computed using the frequency domain formalism. 
\end{abstract}


\maketitle

\section{Introduction}

Current gravitational wave detectors such as LIGO, VIRGO, GEO and TAMA are now starting ``science runs" and the plans for upgrades to these detectors are already in the works. This paper attempts to model a likely occurrence in galactic nuclei i.e. the capture of compact stellar-sized objects by the supermassive black holes that exist in the center of most galaxies. These type of events will generate gravitational waves within the sensitivity band of NASA's space-based LISA project. 

This problem has been treated in the past, quite extensively, using a frequency domain approach to linear black hole perturbation theory \cite{Dav,Detw,Pas,Poi1,Finn,Poi2,Poi3,Poi4,Tana1,Tana2,Tana3,Suz,FT, Hug1,Hug2,Hug3}. A similar calculation in the time domain was only recently presented for the special case of equatorial, circular particle orbits around a rapidly rotating black hole \cite{Ramon}. In this work, we extend the time domain calculation to treat more generic types of orbits, i.e. elliptic and inclined, and compare energy and angular momentum fluxes lost from gravitational wave emission to current frequency domain calculations. We ignore the effects of radiation reaction in this work, with the hope of addressing those in detail sometime in the future. We note that for highly elliptic orbits, both approaches develop problems and only a very detailed study, possibly with the inclusion of radiation reaction effects will lead us to favor one over the other. 

This paper is organized as follows. In the next section we review, briefly, the basics of our time domain approach. Details can be found in the second section of our past work \cite{Ramon}. Then in the following section, we present our results for waveforms and radiative fluxes and compare them with frequency domain results. In the last section, we present our conclusions and emphasize that future work will be based on the inclusion of radiation reaction in our calculations.  

\section{Teukolsky equation in the time domain}

The general approach to this problem  in the time domain, with details regarding numerical implementation, stability, second-order convergence, etc. have been presented in the earlier work on the subject \cite{Ramon}, so we simply summarize some of the essentials here. The Teukolsky equation in Boyer-Lindquist coordinates with the matter-source term, appears below \cite{Teuk}, 
\begin{eqnarray}
&&{}\left [ \frac{(r^2+a^2)^2}{\Delta} - a^2 \sin^2 \theta \right ]
\frac{\partial^2 \psi}{\partial t^2} + \frac{4 M a r}{\Delta}
\frac{\partial^2 \psi}{\partial t
\partial \phi} + \left [ \frac{a^2}{\Delta} - \frac{1}{\sin^2 \theta} \right ]
\frac{\partial^2 \psi}{\partial \phi^2}
- \Delta^{-s} \frac{\partial}{\partial r} \left ( \Delta^{s+1}\frac{\partial
\psi}{\partial r}\right ) \nonumber \\
&&
- \frac{1}{\sin \theta} \frac{\partial}{\partial \theta} \, \left ( \sin
\theta \frac{\partial \psi}{\partial \theta} \right )
-2 s \left [\frac{a (r-M)}{\Delta} + \frac{i
\, \cos \theta}{\sin^2 \theta} \right ] \frac{\partial
\psi}{\partial \varphi} - 2 s \left [ \frac{M(a^2 - r^2)}{\Delta} -
r - i\, a\, \cos \theta \right ] \frac{\partial
\psi}{\partial t} \nonumber \\
&& + \left [ s^2 \cot^2 \theta - s \right ] \psi = 4 \pi (r^2+a^2\cos^2\theta) T
\end{eqnarray}
where $\Delta=r^2-2M r+a^2$. 

The methodology for numerically evolving the homogeneous part of this equation has been presented in detail at various places in the literature \cite{Laguna}. In this work we focus on the matter-source term. In the time domain Teukolsky equation this source term is given by
\begin{equation}
T = 2 \rho^{-4} T_4,
\end{equation}
where
\begin{eqnarray}
&&{}T_4 = (\Delta + 3 \gamma - \gamma^* + 4 \mu + \mu^*)[(\delta^*
- 2 \tau^* + 2 \alpha)T_{nm^*} \nonumber \\ &&{}- (\Delta + 2
\gamma - 2 \gamma^* + \mu^*)T_{m^*m^*}] + (\delta^* - \tau^* +
\beta^* + 3 \alpha + 4 \pi) \nonumber \\ && \times [(\Delta + 2
\gamma + 2 \mu^*)T_{nm^*} - (\delta^* - \tau^* + 2 \beta^* + 2
\alpha) T_{nn}].
\end{eqnarray}
All the greek symbols used above are the ones as defined in the original Teukolsky equation paper \cite{Teuk}. By choosing their values in Boyer-Lindquist coordinates and expanding everything we obtain an explicit but quite complicated form of the source term. We treat the delta functions that appear in the energy-momentum tensor of the particle in orbit by using an approximation in which we substitute the delta function by a very narrow Gaussian function (whose width is only of a few grid points),
\begin{equation}\label{gauss}
\delta(x - x(t))\,\approx\,\frac{1}{\sqrt{2
\pi}\,\sigma}\,\exp\left(\frac{-(x-x(t))^2}{2
\sigma^2}\right) \hspace{0.3 in}{\rm for}\;\sigma \;{\rm small}.
\end{equation}
We handle the deltas in the radial and angular direction through the Gaussian approximation, whereas the $\delta(\phi - \phi(t))$ function we handle analytically. We use code generation software to convert the source term expressions to FORTRAN code that can be used with our Teukolsky equation evolver. We verify that this substitution is adequate by showing convergence of the obtained waveform and radiative flux values as $\sigma\,\rightarrow\,0$.  We also take special care that the Gaussian integrates to unity on the spatial hypersurface. To achieve this with the three-metric of the background black hole spacetime, we normalize the mass-density of the particle by the factor,
\begin{equation}\label{normlz}
  N = \frac{\sqrt{\gamma^{(3)}}}{\sqrt{g^{(3)}}},
\end{equation}
where $\gamma^{(3)}$ is the 3-metric of flat space and $g^{(3)}$
is the 3-metric of a slice of constant Boyer-Lindquist time of the
Kerr spacetime.  

The initial data we take for the fields is zero. Since we are considering the Teukolsky equation with a source this would correspond to the particle ``appearing suddenly'' and that generates an artificial burst of radiation. We handle this by evolving to late times and computing fluxes only when the system has settled down. Sometime in the future, we will examine the implementation of initial data in this context, but at this time we will simply rid the system of this initial burst, by evolving ``long enough''.  

Lastly, we also need a high accuracy general geodesic equation integrator, to obtain the position of the particle at every time step.  The main problem with integrating the geodesic equations is the presence of turning points. At the turning points, since the right-hand sides of the geodesic equations go to zero (see below), the integrators have trouble resolving 
the motion correctly and large amounts of error accumulate. 
\begin{eqnarray}
\Sigma^2\left({dr\over d\tau}\right)^2 && =\left[E(r^2+a^2) - a
L_z\right]^2- \Delta\left[r^2 + (L_z - a
E)^2 + Q\right]\;,\label{eq:rdot} \\
\Sigma^2\left({d\theta\over d\tau}\right)^2 && = Q - \cot^2\theta
L_z^2 -a^2\cos^2\theta(1 -
E^2)\;,\label{eq:thetadot} \\
\Sigma\left({d\phi\over d\tau}\right) && = \csc^2\theta L_z +
aE\left({r^2+a^2\over\Delta} - 1\right) -
{a^2L_z\over\Delta}\;,\label{eq:phidot} \\
\Sigma\left({dt\over d\tau}\right) && =
E\left[{(r^2+a^2)^2\over\Delta} - a^2\sin^2\theta\right] +
aL_z\left(1 - {r^2+a^2\over\Delta}\right)\;.\label{eq:tdot}
\label{eq:geodesiceqns}
\end{eqnarray}
Fortunately, there is such a public domain integrator, called ``Geod'' written by Scott Hughes \cite{Geod}. In Geod, this problem is solved by transforming the equations to orbital parameters that do not have turning points. More information can be found about this approach in the documentation of this software package. We call this geodesic equation integrator from within our Teukolsky code to update the particle's position at every time step.

\section{Results}

\begin{table}
  \centering
     \begin{tabular}{|c||c||c|} \hline
   {\bf m mode} & {\bf Time domain energy flux} & {\bf Frequency domain flux }\cr \hline \hline
    $1$ &      $2.9\times10^{-10}$ &  $2.8\times10^{-10}$ \cr \hline
    $2$ &      $1.4\times10^{-07}$ &  $1.8\times10^{-07}$ \cr \hline
    $3$ &      $4.2\times10^{-08}$ &  $5.3\times10^{-08}$ \cr \hline   
   \end{tabular}
 \caption{Comparisons of gravitational wave energy fluxes detected at infinity using the frequency domain solution  with the results measured numerically  at $r = 100 M$ while evolving the Teukolsky equation in the time domain as discussed here. The particle here is in an elliptic, equatorial orbit of semi-latus rectum $2.0\, r_{isco}$ and eccentricity of $0.5$ about a black hole with Kerr parameter  $a/M = 0.9$ and $\mu/M=0.01$.}\label{comp1}
\end{table}

In this section, we present sample waveforms and compute radiative fluxes from numerical evolutions of the Teukolsky equation with matter-source term as discussed in the previous section. As detailed before \cite{Ramon}, the numerical implementation of this work is in the form of a set of 2+1 dimensional linear PDE's that uses the Lax-Wendroff evolution scheme. Also discussed therein is the convergence and stability of the entire approach, therefore that study shall not be presented here.  

We calculate the energy and angular momentum fluxes emitted due to gravity wave emission from the binary for a sample equatorial, elliptic orbit and show close agreement with current results of high accuracy frequency domain calculations \cite{Dan}. The code monitors the value of the Teukolsky function at a set radial location in the grid far from the horizon 
and computes the energy and angular momentum fluxes in gravitational radiation according to the formulae \cite{ djdt1} :
\begin{eqnarray}
\frac{{\rm d} E}{{\rm d} t}&=&\lim_{r\to\infty}\left\{\frac{1}{4
\pi r^6} \int_{\Omega} {\rm d}\Omega \left| \int_{-\infty}^{t}
\,{\rm
d}\tilde{t}\,\psi(\tilde{t},r,\theta,\varphi)\right|^{2}\right\} \\
\frac{{\rm d} L_z}{{\rm d}t}&=&-\lim_{r\to\infty}\left\{\frac{1}{4
\pi r^6}\mathbf{Re}\left[\int_{\Omega} {\rm d}\Omega
\left(\partial_{\varphi}\,\int_{-\infty}^{t} \,{\rm
d}\tilde{t}\,\psi(\tilde{t},r,\theta,\varphi)\right)\right.\right.\nonumber\\
&&\left.\left.\times\left(\int_{-\infty}^{t} \,{\rm d}t'\,
\int_{-\infty}^{t'} \,{\rm
d}\tilde{t}\,\bar{\psi}(\tilde{t},r,\theta,\varphi)\right)\right]\right\}\;
\;\;\;\;\, {\rm d}\Omega=\sin\theta{\rm d}\theta{\rm d}\varphi.
\label{jfl}
\end{eqnarray}

In Tables \ref{comp1} and  \ref{comp2} we present representative comparisons with frequency domain calculations for equatorial, elliptic orbits. The level of agreement between the two approaches is about the same (within  $25\%$) over a wide range of eccentricities ($0.0$ through $0.8$). It is interesting to note both frequency domain approaches and time domain approaches begin to have computational issues for orbits of high eccentricity (above $0.8$) and also for hyperbolic orbits. In the frequency domain,  where the method is based on variable separating the Teukolsky equation the need to compute for a large number of $k$ and $\ell$ modes for a given accuracy, makes it computationally very demanding (see Figure 1).  In the time domain case, since highly elliptic orbits have large orbital periods, we need to evolve for very long times (several tens of thousands times the mass of the hole) in order to compute average radiation fluxes (see Figure 3).  Therefore it is not clear at this point, which of the two approaches is better suited to address the general problem of extreme mass ratio inspiral. However, this question will be revisited, once more work is done on the inclusion of  radiation reaction and a more detailed comparison study is undertaken. 

We also include a representative waveform in Figure 2 from the evolution of a particle in an equatorial, elliptic orbit. It shows the real part of the $m=3$ mode of the Teukolsky function, and as mentioned before, the  initial burst of radiation is unphysical, since it corresponds to the particle ``appearing out of nowhere''. It is only after the evolution has settled down to a time-periodic pattern, do we compute average fluxes. In Figure 3 we show the variation of energy fluxes with time, from the evolution of a particle in an equatorial, elliptic orbit and also in an inclined, circular orbit. Its is interesting to note that if we keep the semi-latus rectum fixed, and study the energy-fluxes as they change with eccentricity for elliptic-equatorial orbits, they increase at first, but then begin to drop after reaching a maximum at an eccentricity of about $0.5$. It turns out that this variation can be understood easily by studying the same in the context of a binary Newtownian system where this turn-over point can be computed to be about $0.47$ \cite{sam}.

Lastly, we suggest here an alternate approximate approach to treating delta functions on a spatial grid. This approach is significantly more efficient (more than two orders of magnitude faster in a numerical implementation) and yields significantly more accurate results (in terms of their agreement with frequency domain) compared with the above mentioned narrow Gaussian approach. One can think of this as a limiting case of the Gaussian particle approach, where the width of the Gaussian is made so narrow that it is comparable to the numerical grid separation. In this approach, we implement a ``discrete'' version of the delta function on the spatial grid. More specifically, we define a discrete-delta function as being unity at the appropriate grid point and zero everywhere else. We then define the derivatives of the delta function as the result of applying the usual finite-difference derivative stencils to the above mentioned discrete-delta function. Of course, the discrete-delta needs to be normalized suitably, i.e. scaled by a factor such that its integral on the spatial grid is unity. Note that such a delta function satisfies the usual delta function properties, such as $f(x)\delta'(x)=-f'(x)\delta(x)$, on the discrete grid. Care has to be taken to study convergence of the code using such an approach, however, just the efficiency and accuracy obtained by using this simple idea, makes it worth mentioning and investigating further. 

\section{Conclusions}

We can now treat different types of particle orbits in the context of extreme mass ratio inspiral, using black hole perturbation theory in the time domain. We get good agreement in comparisons with current frequency domain results for energy and angular momentum radiative fluxes.  At this time, it is not clear which of these two approaches is better suited to the generic case and will provide more accurate results, more efficiently.  

In future work, we hope to be able to further widen the scope of these calculations by incorporating the case of a spinning particle in orbit about a Kerr hole and we also expect to investigate methods to include radiation reaction. The first approach will be to correct the orbit trajectory using energy-balance arguments, as for instance in \cite{Hug1}. But, later other proposals for the inclusion of radiation reaction forces will be incorporated. 

\begin{table}
  \centering
     \begin{tabular}{|c||c||c|} \hline
   {\bf m mode} & {\bf Time domain ang. mom. flux} & {\bf Frequency domain flux }\cr \hline \hline
    $1$ &      $2.4\times10^{-09}$ &  $2.3\times10^{-09}$ \cr \hline
    $2$ &      $1.0\times10^{-06}$ &  $1.4\times10^{-06}$ \cr \hline
    $3$ &      $3.0\times10^{-07}$ &  $4.0\times10^{-07}$ \cr \hline   
   \end{tabular}
 \caption{Comparisons of gravitational wave angular momentum fluxes detected at infinity using the frequency domain solution  with the results measured numerically  at $r = 100 M$ while evolving the Teukolsky equation in the time domain as discussed here. The particle here is in a elliptic, equatorial orbit of semi-latus rectum $2.0\, r_{isco}$ and eccentricity of $0.5$ about a black hole with Kerr parameter  $a/M = 0.9$ and $\mu/M=0.01$.}\label{comp2}
\end{table}

\acknowledgments

We express our gratefulness to Daniel Kennefick for working very closely with us and generating results for comparisons using his frequency domain codes. We also thank Jorge Pullin, Pablo Laguna, Ramon Lopez-Aleman, Eric Poisson, Karl Martel, Lee Samuel Finn and Scott Hughes for input and discussions all along, relating to this research. This work was supported by NSF grant number PHY-0140236. We also thank the Long Island University for its continued research support. Some of the numerical simulations were performed at the Boston University's Scientific Computing and Visualization Group. We thank them for allocating us supercomputing time on their IBM p690 server.

\begin{figure}
\epsfysize=80mm \epsfbox{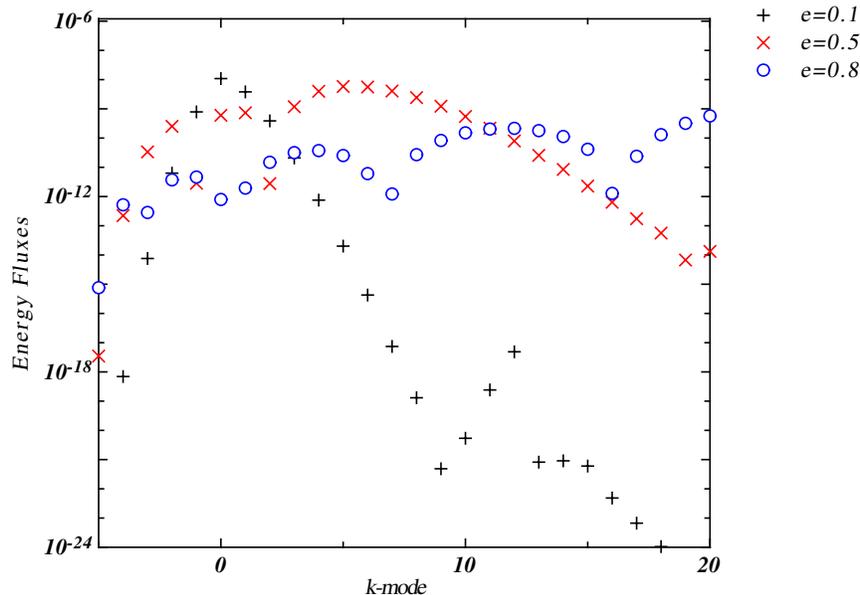} \caption{Energy fluxes computed from the evolution of the $m=3$ mode of the Teukolsky function from a particle ($\mu/M=0.01$) orbiting a rapidly rotating Kerr hole of $a = 0.9 M$. These results were obtained from a frequency domain computation, and  are plotted against the $k$ modes.  All the orbits are equatorial and have a semi-latus rectum of  $2\,r_{isco}$. Note that as the orbit's eccentricity increases, results from more $k$ modes are necessary, increasing the computational needs of the problem. }
\end{figure}

\begin{figure}
\epsfysize=80mm \epsfbox{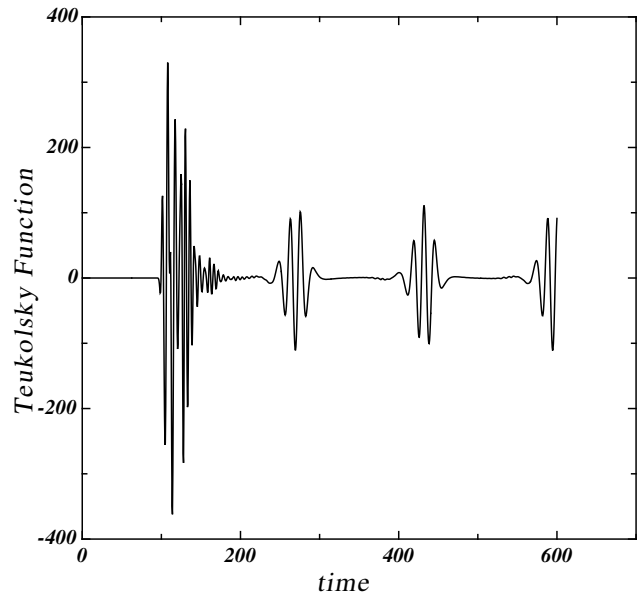} \caption{Representative data showing the evolution of the real part of the Teukolsky function for the $m=3$ mode of a particle ($\mu/M=0.01$) orbiting a rapidly rotating Kerr hole of $a = 0.9 M$ in an elliptic orbit of semi-latus rectum $2\, r_{isco}$ 
($r_{isco}=2.32 M$) and eccentricity of $0.5$ at an observation point $r/M=100$, $\theta=\pi/2$.}
\end{figure}

\begin{figure}
\epsfysize=120mm \epsfbox{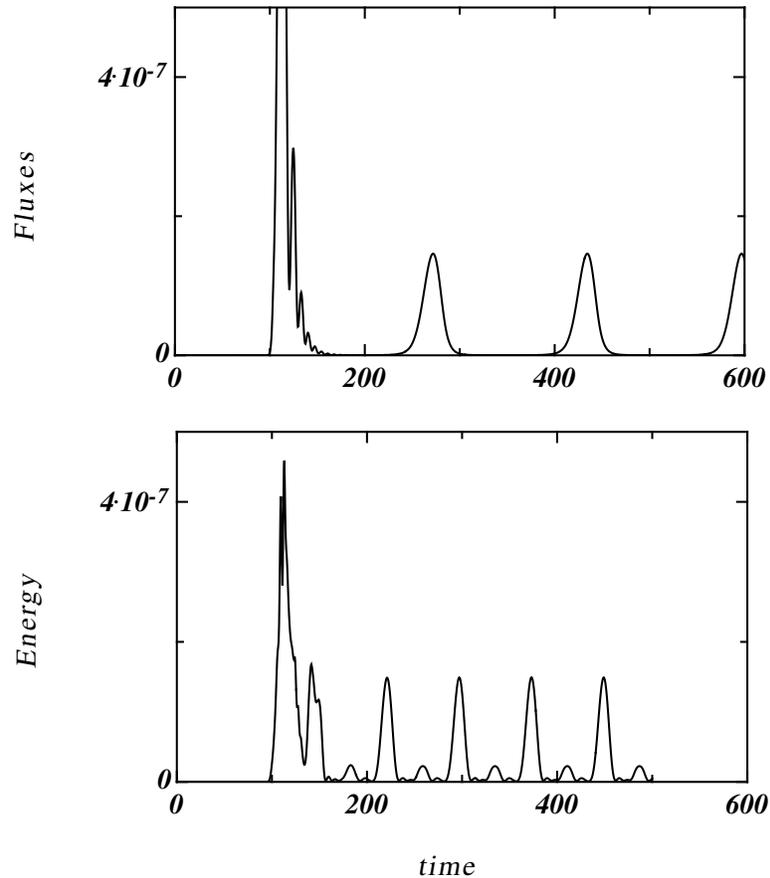} \caption{Energy fluxes computed from the evolution 
of the $m=3$ mode of the Teukolsky function from a particle ($\mu/M=0.01$) orbiting a rapidly rotating Kerr hole of $a = 0.9 M$. The top plot is from the particle in an equatorial, elliptic orbit of semi-latus rectum $2\, r_{isco}$ ($r_{isco}=2.32 M$) and eccentricity of $0.5$ at an observation point $r/M=100$. It should be clear from this plot, that as eccentricity increases, longer evolutions are needed to compute average fluxes. The bottom plot is from the particle in a circular, inclined orbit with radius $2\,r_{isco}$ and initial inclination angle of $\pi/4$ radians. }
\end{figure}

\end{document}